\documentclass[12pt]{iopart} 

\usepackage[dvips]{epsfig}
\usepackage{iopams}
\usepackage{amstext}
\renewcommand{\vec}[1]{{\bi{#1}}}

\begin{document}

\jl{3} 

\title[Ferromagnetism in the Periodic Anderson Model]{$f$-Band
  Ferromagnetism in the Periodic Anderson Model --  a Modified Alloy
  Analogy}

\author{G G Reddy\dag, D Meyer\ddag, S Schwieger\ddag, A Ramakanth\dag
  and W Nolting\ddag} 
\address{\dag Department of Physics, Kakatiya University,
  Warangal-506009, India}
\address{\ddag Institut f{\"u}r Physik, Humboldt-Universit{\"a}t zu
  Berlin, D-10115 Berlin, Germany}

\begin{abstract}
We introduce a new aproximation scheme for the periodic Anderson model
(PAM). The modified alloy approximation represents an optimum alloy
approximation for the strong coupling limit, which can be solved within
the CPA-formalism. Zero-temperature and finite-temperature phase
diagrams are presented for the PAM in the intermediate-valence
regime. The diversity of magnetic properties accessible by
variation of the system parameters
can be studied by means of quasiparticle densities of
states: The conduction band couples either ferro- or
antiferromagneticaly to the $f$-levels. A finite hybridization is
a necessary precondition for ferromagnetism. However, too strong
hybridization generally 
suppresses ferromagnetism, but can for certain system parameters also
lead to a semi-metallic state
with unusual magnetic properties. By comparing with the spectral density
approximation, the influence of quasiparticle damping can be examined.
\end{abstract}


\section{Introduction}
\label{sec:intro}
Correlated electron systems have come to occupy the centre stage in both
theory and experiment in condensed matter physics. There is a general
consensus that strong electron correlations play a decisive role in a
variety of phenomena such as magnetism, heavy fermions, high-temperature
superconductivity, colossal magneto-resistance etc. The real systems that
display these phenomena are transition metals (3d), alloys and compounds
as well as rare earth (4f) systems (metals, insulators), in particular
cuprates, manganites, Ce compounds etc.. There is intense activity both
theoretically and experimentally to isolate the essential interactions
responsible for these phenomena. In this process, several model systems
are being studied which are expected to at least qualitatively mimic the
actual systems in some region or the other of the parameter space. One
of the most widely used models which brings out the role of electron
correlations while also taking into account the interplay of two
different types of electrons, one highly localized and the other quasi
free, is the well-known Anderson model \cite{hewson}. One distinguishes
the "single impurity Anderson model"(SIAM), if the system of
uncorrelated conduction electrons hybridizes with a single localized
state, and the "periodic Anderson model" (PAM), if the hybridization
takes place with a periodic lattice of localized states. As far as the
SIAM is concerned many exact results have been obtained, e. g. by using
Bethe ansatz \cite{TW83} or renormalization group theory
\cite{Wil75}. Very recently the SIAM has gained a new upsurge in
connection with the "dynamical mean field theory"
\cite{MV89,PJF95,GKKR96} that exploits the
fact that in infinite lattice dimensions theoretical lattice models like
the Hubbard model can be mapped onto single-impurity models such as the
SIAM. Numerically essentially exact methods such as the exact
diagonalization of small systems, quantum Monte Carlo calculations or
numerical renormalization group theory
\cite{GKKR96,BHP98} 
have provided further insight into the physics of the model. However,
since these numerical methods are bound to certain limitations, reliable
analytical approaches remain to be required \cite{MWPN99}.

The situation is less satisfactory in the
case of the PAM. Even though there exists a number of approximate schemes
for different parameter constellations, referring to either of the
so-called heavy fermion, intermediate valence and Kondo regimes, there
is still need for further improvement and extension. The matter acquires
additional urgency in view of the fact that for the parameter space
corresponding to strongly correlated low-lying localized states the PAM can
be mapped \cite{SW66} on to the Kondo lattice model \cite{Don77} which
is of great
current interest, e. g. with respect to the extraordinary physical
properties of the manganites \cite{Ram97}.  

Recently we proposed a new
approximation scheme \cite{MNRR98} which is based on a mapping of the
PAM on to 
an effective Hubbard model. The parameters of the Hubbard model are such
that they correspond to the strong coupling limit. Therefore we could
exploit a reliable approximate theory, trustworthy in the strong
coupling limit. Such a theory is the "spectral density approach" (SDA)
\cite{GN88,NB89,HN97a}. Its main advantages are the physically simple
concept and
the non-perturbative character. In this scheme we have studied the
magnetic T = 0-phase diagram of the PAM as well as its finite
temperature magnetic properties. The SDA, however, shows up a principal
limitation concerning quasiparticle damping. By ansatz the SDA
self-energy is a real quantity, thus neglecting from the very beginning
effects due to finite lifetimes of the quasiparticles. How quasiparticle
damping influences the magnetic stability in the PAM is an interesting
and important question having been left open by our previous theory
\cite{MNRR98}. It is the main impetus for the present study to close
this gap.   

A
conceptionally simple method that is able to provide complex
self-energes is the "coherent potential approximation" (CPA)
\cite{VKE68}. In 
order to perform CPA on the PAM, we have to fix up an alloy analogy. This
requires the determination of the
energy levels and the concentrations of the components of the fictitious
alloy. Within the conventional alloy analogy treatment (AA) due to
Hubbard \cite{Hub64b} these are taken by referring to the atomic
limit. For an application to the PAM, see e.\ g.\
\cite{LC79b,Czy86}. However, 
this choice is by no means predetermined. On the contrary it can be
shown that it is indeed not the best ansatz. One knows that CPA becomes
an exact procedure in infinite lattice dimensions ($d =\infty$), where
the 
inherent single-site aspect of CPA is rigorous \cite{VV92}. However, the
CPA-solution of the AA for the $d=\infty$-Hubbard model violates the
exactly known strong coupling behaviour \cite{HL67} and also does not
reproduce the weak coupling results of second order perturbation theory
\cite{BJ90,PN97}. Furthermore it contradicts exact high-energy expansions
\cite{PHWN98}. So we have to conclude that the conventional AA, which
starts from the atomic limit 
solution, is not the most convenient alloy analogy. In this paper we
therefore derive the "modified
alloy analogy" (MAA), which was recently introduced for the Hubbard
model\cite{HN96,PHWN98}, for the PAM.
This method substantially improves the AA ansatz by deriving the proper
alloy analogy from exact high energy expansions of the single-electron
Green function and the self-energy, respectively. This procedure
guarantees the correct strong coupling behaviour relevant for the PAM.
On the way of deriving the MAA, the above-mentioned SDA can
easily be re-derived without the necessity of introducing the
effective Hubbard model as was done before\cite{MNRR98}.
It turns out that the "atomic levels" of
fictitious alloy constituents are nothing else but the SDA-quasiparticle
energies in the zero-bandwidth limit. Similarly the "concentrations"
agree with the SDA-spectral weights in this limit. The MAA therewith
promises to keep the advantages of the SDA while simultaneously
improving the method by incorporating quasiparticle damping. The
SDA-energies and spectral weights contain non-trivial thermodynamic
expectation values which have to be determined self-consistently. By
them the itineracy of the $-\sigma$-electrons, which define the fictitious
alloy for the propagating $\sigma$-electrons, comes into play, at least
to a 
certain degree. The neglect of this itineracy is a well-known
shortcoming of the conventional AA. 

The paper is organized as
follows. In the next section the PAM and its many-body problem is
formulated. An approximate solution by use of the MAA
scheme is then developed in section 3. Section 4 is devoted to a
presentation and discussion of the results concerning the magnetic
properties of the PAM. The paper ends with some concluding remarks.

\section{Model Hamiltonian and Its Many-Body Problem}
Starting point is the periodic Anderson Hamiltonian
\begin{eqnarray}
  \label{hamiltonian}
     H =&\sum_{ij\sigma} (T_{ij}-\mu) s_{i\sigma}^{\dagger}s_{j\sigma} +
    \sum_{i,\sigma} (e_{\rm f}-\mu) f_{i\sigma}^{\dagger}f_{i\sigma} +
    \\&+V \sum_{i,\sigma} (f_{i\sigma}^{\dagger}s_{i\sigma} +
    s_{i\sigma}^{\dagger}f_{i\sigma} ) + \frac{1}{2} U \sum_{i,\sigma}
    n_{i\sigma}^{(f)}n_{i-\sigma}^{(f)}  \label{hhh}
\end{eqnarray}
$s_{i\sigma}$ ($f_{i\sigma}$)and $s_{i\sigma}^{\dagger}$
($f_{i\sigma}^{\dagger}$) are, respectively, the annihilation and the
creation operators for
an electron in a non-degenerate conduction band (localized $f$-state), and
$n^{(f)}_{i\sigma}=f^{\dagger}_{i\sigma}f_{i\sigma}$ is the
spin-dependent occupation
number operator for the f-state. The index i
refers to 
the respective lattice site, $\sigma=\{\uparrow,\downarrow\}$ is the
spin projection. The
hopping integral $T_{ij}$,
\begin{equation}
 \label{hopping-integral}
  T_{ij}=\frac{1}{N}\sum_{\vec{k}} e^{-i\vec{k}
    \cdot(\vec{R}_i-\vec{R}_j)}\epsilon(\vec{k}) 
\end{equation}
describes the propagation of a band electron from site $R_j$ to site
$R_i$. $\epsilon(\vec{k})$ is the free Bloch energy, while $e_{\rm f}$
denotes 
the position of the
non-degenerate $f$-level. We choose the energy zero to coincide with the
centre of gravity of the unperturbed conduction band: 
\begin{equation}\label{tiij}
 T_{ii}=\frac{1}{N}\sum_{\vec{k}}\epsilon(\vec k)\stackrel{!}{=}0
\end{equation}
$U$ is the intra-atomic Coulomb repulsion between $f$-electrons. The
hybridization $V$ is taken as a real and $\vec{k}$-independent local matrix
element, $\mu$ is the chemical potential.

Let us start with the retarded single-$s$($f$) electron Zubarev Green's
function
\begin{eqnarray}
  \label{eq:greenfunction}
  G_{ij\sigma}^{(f)}(E) = \langle\!\langle f_{i\sigma} ;
  f_{j\sigma}^{\dagger} \rangle\!\rangle;\quad  G_{ij\sigma}^{(s)}(E) =
  \langle\!\langle s_{i\sigma} ; 
  s_{j\sigma}^{\dagger} \rangle\!\rangle\\
G_{\vec k \sigma}^{(f,s)}=\frac{1}{N}\sum_{\vec{k}}
e^{i\vec{k}\cdot(\vec{R}_i-\vec{R}_j)} G_{ij\sigma}^{(f,s)}(E)\label{ggggg}
\end{eqnarray}
The equations of motion are easily derived and formally solved,
\begin{eqnarray}
  \label{eq:greenfunction2s}
  G_{\vec{k}\sigma}^{(s)}(E) =\hbar
  \frac{E-(e_{\rm f}-\mu)-\Sigma_{\vec{k}\sigma}(E)}{
    (E-(e_{\rm f}-\mu)-\Sigma_{\vec{k}\sigma}(E) ) (E-(\epsilon(\vec{k})-\mu))
    -V^2}\\
  \label{eq:greenfunction2f}
    G_{\vec{k}\sigma}^{(f)}(E) = 
    \frac{\hbar}
    {E-(e_{\rm f}-\mu)- \frac{V^2}{E-(\epsilon(\vec{k})-\mu)}
      -\Sigma_{\vec{k}\sigma}(E)}  
\end{eqnarray}
where the self-energy $\Sigma_{\vec{k}\sigma}(E)$ has been introduced via
\begin{equation}
  \label{eq:sigma}
  \Sigma_{\vec{k}\sigma}(E) G_{\vec{k}\sigma}^{(f)}(E) = U \frac{1}{N}
  \sum_{\vec{p},\vec{q}}\langle\!\langle f_{\vec{p}\sigma}^{\dagger}
  f_{\vec{q}-\sigma} f_{\vec{p}+\vec{k}-\vec{q}\sigma};
  f_{\vec{k}\sigma}^{\dagger} \rangle\!\rangle
\end{equation}
Obviously the explicitly problem is solved as soon as we have found a
solution for 
the self-energy. 
The $f$- and $s$-quasiparticle densities of states (QDOS) are given by
\begin{eqnarray}
  \label{rhos}
  \rho_{\sigma}^{(s)}(E)=- \frac{1}{\pi \hbar N} \sum_{\vec{k}} \Im
  G_{\vec{k}\sigma}^{(s)}(E-\mu+\rmi 0^+)\\
  \label{rhof}
  \rho_{\sigma}^{(f)}(E)=- \frac{1}{\pi \hbar N} \sum_{\vec{k}} \Im
  G_{\vec{k}\sigma}^{(f)}(E-\mu+\rmi 0^+)
\end{eqnarray}
and determine the spin-dependent average occupation number
$n^{(s,f)}_{\sigma}$, 
which we need to construct the magnetic phase diagram of the PAM:
\begin{eqnarray}
  \label{ns}
  n_{\sigma}^{(s)}=\langle s_{i\sigma}^{\dagger}s_{i\sigma}\rangle =
  \int_{-\infty}^{+\infty} dE f_-(E) \rho_{\sigma}^{(s)}(E)\\
  \label{nf}
  n_{\sigma}^{(f)}=\langle f_{i\sigma}^{\dagger}f_{i\sigma}\rangle =
  \int_{-\infty}^{+\infty} dE f_-(E) \rho_{\sigma}^{(f)}(E)
\end{eqnarray}

\section{Modified alloy analogy}
\label{sec:maa}
A standard method to solve many-body problems as that posed by the
PAM Hamiltonian~(\ref{hamiltonian}) is the CPA \cite{VKE68}. 
As a single-site approximation the resulting $f$-electron self-energy
will be local, i.\ e.\ $\vec{k}$-independent. If we apply
this method here we have to solve the following
equation self-consistently
\begin{eqnarray}
  \label{cpa}
  0=\sum_{p=1}^{n} x_{p\sigma}\frac{E_{p\sigma}-\Sigma_{\sigma}(E)-e_{\rm f}}
  {1-\frac{1}{\hbar}G_{ii\sigma}^{(f)}(E) (E-\Sigma_{\sigma}(E)-e_{\rm f})}\\
  \label{gii}
  G_{ii\sigma}^{(f)}(E)=\frac{1}{N}\sum_{\vec{k}}
  G_{\vec{k}\sigma}^{(f)}(E); \qquad
  \Sigma_{\sigma}(E)=\frac{1}{N}\sum_{\vec{k}}
  \Sigma_{\vec{k}\sigma}(E) 
\end{eqnarray}
The solution of (\ref{cpa}) needs to fix
the alloy 
analogy, i.\ e.\ the "atomic levels" $E_{p\sigma}$ and the
"concentrations" $x_{p\sigma}$ of the
$n$ constituents of the fictitious alloy. Since the dominant features of
the correlated $f$-level in the PAM are the two charge excitations
separated by $U$, the number of ``alloyed'' components $n$ is set to two.
The conventional alloy analogy
(AA) uses the $V=0$ limit (``atomic'' limit) of the PAM to determine
$E_{p\sigma}$ and $x_{p\sigma}$\cite{LC79b,Czy86}:
\begin{eqnarray}
  \label{aa}
  E_{1\sigma}^{(AA)}=e_{\rm f};&   x_{1\sigma}^{(AA)}=n_{-\sigma}^{(f)}\\
  E_{2\sigma}^{(AA)}= e_{\rm f}+U; \qquad& x_{2\sigma}^{(AA)}=
  1-n_{-\sigma}^{(f)}\nonumber 
\end{eqnarray}
The result violates the weak as well as the strong coupling behaviour
and as for the Hubbard model prohibits spontaneous magnetism
\cite{LC79b,FE73,SD75,PHWN98}. In
particular the strong coupling behaviour appears to be crucial with
respect to ferromagnetism. On the other hand, the choice (\ref{aa}) is not
at all predetermined. We propose another way to find out the "best alloy
analogy". Correct strong coupling behaviour is guaranteed by fulfilling
the high-energy expansion of relevant Green functions and self-energies
\cite{PHWN98}. Decisive ingredients for proper high-energy expansions are the
local spectral moments
\begin{eqnarray}
  M_{\sigma}^{(n)}=\frac{1}{N}\sum_{\vec{k}}
  M_{\vec{k}\sigma}^{(n)};\quad n=0,1,2,\dots\nonumber\\
  \label{moments}
  M_{\vec{k}\sigma}^{(n)}=\frac{1}{\hbar}\int dE E^n S_{\vec{k}\sigma}(E)
\end{eqnarray}
$S_{\vec{k}\sigma}(E)$ is the $f$-electron spectral density:
\begin{equation}
  \label{eq:spec_dens}
  S_{\vec{k}\sigma}(E)=- \frac{1}{\pi} \Im
  G_{\vec{k}\sigma}^{(f)}(E+\rmi 0^+)
\end{equation}
The moments can be calculated independently of the required spectral density:
\begin{equation}
  \label{eq:moments}
  M_{\vec{k}\sigma}^{(n)}=\langle  [
  \underbrace{[\dots[f_{\vec{k}\sigma},H]_-,\dots,H]_-}_{\text{$n$-fold
      commutator}}, 
  f_{\vec{k}\sigma}^{\dagger} ]_+\rangle 
\end{equation}
$[\dots,\dots]_-$ denotes the
commutator and $[\dots,\dots]_+$ the
anti-commutator. For the Green function (\ref{eq:greenfunction}) we can write 
\begin{equation}
  \label{eq:expansionG}
  G_{\vec{k}\sigma}^{(f)}(E)=\int dE' \frac{S_{\vec{k}\sigma}(E')} {E-E'}
  = \hbar \sum_{n=0}^{\infty} \frac{M_{\vec{k}\sigma}^{(n)}}{E^{n+1}}
\end{equation}
For solving the CPA-equation~(\ref{cpa}) we only need the local
spectral moments 
which are found to be
\begin{eqnarray}
  \label{eq:moments2}
  M_{\sigma}^{(0)} =& 1\nonumber \\
  M_{\sigma}^{(1)} =& e_{\rm f} + U n_{-\sigma}^{(f)}\nonumber\\
  M_{\sigma}^{(2)} =& e_{\rm f}^2 + 2 e_{\rm f} U n_{-\sigma}^{(f)} +U^2
  n_{-\sigma}^{(f)} +V^2\\
  M_{\sigma}^{(3)}  =& e_{\rm f}^3 + 3 e_{\rm f}^2 U n_{-\sigma}^{(f)} +
  U^2 e_{\rm f} (2 
  n_{-\sigma}^{(f)} + n_{-\sigma}^{(f)^2}) + U^3 n_{-\sigma}^{(f)}+
  \nonumber\\&+   
  V^2(2 e_{\rm f} +2 U n_{-\sigma}^{(f)} + T_{ii}) +U^2 n_{-\sigma}^{(f)}
  (1-n_{-\sigma}^{(f)}) B_{-\sigma} \nonumber
\end{eqnarray}
From (\ref{eq:greenfunction2f}) we get the corresponding expansion for
the self-energy: 
\begin{equation}
  \label{eq:expansionS}
  \Sigma_{\vec{k}\sigma}(E)=\sum_{n=0}^{\infty}
  \frac{C_{\vec{k}\sigma}^{(n)}} {E^n};\qquad 
  \Sigma_{\sigma}(E)=\frac{1}{N}\sum_{\vec{k}}   \Sigma_{\vec{k}\sigma}(E)
\end{equation}
with the local coefficients:
\begin{eqnarray}
  C_{\sigma}^{(0)} &= U n_{-\sigma}^{(f)}\nonumber\\
  \label{S_coeff}
  C_{\sigma}^{(1)} &= U^2 n_{-\sigma}^{(f)} (1-n_{-\sigma}^{(f)})\\
  C_{\sigma}^{(2)} &= U^2 n_{-\sigma}^{(f)} (1-n_{-\sigma}^{(f)})
  (B_{-\sigma} +U(1-n_{-\sigma}^{(f)}))
  \nonumber
\end{eqnarray}
Surprisingly the hybridization $V$ does not explicitely appear in the
$C_{\sigma}^{(n)}$. The contributions via the moments
(\ref{eq:moments2}) are exactly cancelled by those from the term
$\frac{V^2}{E-(\epsilon(\vec{k})-\mu)}$ in (\ref{eq:greenfunction2f}).

By use of (\ref{eq:expansionG}) and (\ref{eq:expansionS}) we now expand
the CPA-equation (\ref{cpa}) with
respect to powers of $\frac{1}{E}$. Comparison of the
coefficients of the $\frac{1}{E^n}$ terms up to $n=3$ yields the
following set of four equations for the four unknown quantities
$E_{(1,2)\sigma}$ and $x_{(1,2)\sigma}$:

\begin{eqnarray}
  \sum_{p=1}^{2} x_{p\sigma} &=1\nonumber\\
  \label{xxcpa}
  \sum_{p=1}^{2} x_{p\sigma} (E_{p\sigma}-e_{\rm f}) &= Un_{-\sigma}^{(f)}\\
  \sum_{p=1}^{2} x_{p\sigma} (E_{p\sigma}-e_{\rm f})^2 &=
  U^2n_{-\sigma}^{(f)}\nonumber\\ 
  \sum_{p=1}^{2} x_{p\sigma} (E_{p\sigma}-e_{\rm f})^3 &=
      U^3n_{-\sigma}^{(f)}+ U^2n_{-\sigma}^{(f)}(1-n_{-\sigma}^{(f)})
      (B_{-\sigma}-e_{\rm f}) \nonumber
\end{eqnarray}
We now use eqs. (\ref{xxcpa}) to fix the "optimum" alloy
analogy. After simple manipulations we get
\begin{eqnarray}
  \label{maa}
  \fl E_{1,2\sigma}=\frac{1}{2} [ B_{-\sigma}+U+e_{\rm f} \pm
  \sqrt{(B_{-\sigma}+U-e_{\rm f})^2+ 4 U n_{-\sigma}^{(f)}
    (e_{\rm f}-B_{-\sigma})}]\\
  x_{1\sigma}=\frac{E_{2\sigma}-e_{\rm f}-Un_{-\sigma}^{(f)}}{E_{2\sigma} -
    E_{1\sigma}} = 1-x_{2\sigma}\nonumber
\end{eqnarray}
The decisive point for the ongoing procedure is the "higher" correlation
function $B_{-\sigma}$:
\begin{equation}
  \label{eq:bandshift}
  n_{-\sigma}^{(f)}(1-n_{-\sigma}^{(f)})(B_{-\sigma}-e_{\rm f}) = V \langle
  f_{i-\sigma}^{(\dagger)} s_{i-\sigma}(2n_{i\sigma}^{(f)}-1)\rangle
\end{equation}
It has been elaborated in previous works \cite{GN88,NB89,HN97a,PHWN98}
that its analogue 
has a decisive influence on the stability at spontaneous magnetism in
the Hubbard model. We believe, it plays a similarly important role for
ferromagnetism in the PAM\cite{MNRR98,MN00a}.
In spite of the fact that it is a
"higher" correlation function it can rigorously be expressed by
single-electron terms \cite{GN88}:
\begin{eqnarray}
  \label{eq:bandshift2}
  \fl n_{-\sigma}^{(f)}(1-n_{-\sigma}^{(f)})(B_{-\sigma}-e_{\rm f}) = -
  \frac{1}{\pi \hbar} \Im \int_{-\infty}^{+\infty} dE f_-(E)
  \left(\frac{2}{U} \Sigma_{\sigma}(E) -1\right)\\
  \times \left(\left(E -(e_{\rm f}-\mu)-
  \Sigma_{\sigma}(E)\right) G_{ii\sigma}^{(f)}(E) -\hbar\right)
\nonumber 
\end{eqnarray}

With the fictitious alloy (\ref{maa}), we enter the CPA-equation (\ref{cpa}).
The theory is now complete. The equations (\ref{cpa}),
(\ref{gii}), (\ref{eq:greenfunction2f}),
(\ref{eq:bandshift2}) and (\ref{nf}),
together with (\ref{maa}) build a closed system of equations
which can be solved self-consistently for the self-energy
$\Sigma_{\sigma}(E)$. Note that 
the only $\vec{k}$-dependence comes into play by
(\ref{eq:greenfunction2f}) via the formal
solution of the equation of motion. The
$\vec{k}$-dependence is
therefore strictly an $\epsilon(\vec{k})$-dependence, so that all
k-summations
can be replaced by simpler energy-integration over the "free" Bloch
density of states $\rho_0(E)=1/N \sum_{\vec{k}}
\delta(E-\epsilon(\vec{k}))$ which has to be considered as a model
parameter:
\begin{equation}
  \label{eq:gf_final}
  G_{ii\sigma}^{(f)}(E)=\hbar \int_{-\infty}^{+\infty} dx
  \frac{\rho_0(x)}{E-(e_{\rm f}-\mu)-\frac{V^2}{E-(x-\mu)}-\Sigma_{\sigma}(E)}
\end{equation}

Let us comment on the modified alloy analogy (\ref{maa}) and its
differences to the conventional AA~(\ref{aa}): 
Within the AA, the $\sigma$-$f$ electron is propagating through a
fictitious alloy where one component is represented by
lattice sites with no $-\sigma$-electron present, and the other
component by lattices sites which are occupied by one $-\sigma$
electron. This approach therefore completely neglects the hybridization
between the $f$-levels and the conduction band. For calculating the
self-energy, the $-\sigma$ $f$-electrons are frozen, any
exchange with the conduction band is eliminated. This excludes the
possiblity of magnetic order in the system. How can this drawback be
circumvented? It is clear that the PAM will always be considered to
belong to the strong coupling regime (large $U$). Two well separated
excitation peaks will therefore be expected in the
$f$-electron quasiparticle density of states. The above-described ansatz
of fitting the positions and 
weights of these to an exact $\frac{1}{E}$-expansion of the Green
function seems the most plausible one since these charge excitation
peaks themselves are
high-energy features. It is noteworthy that all thereby induced
differences between the 
MAA and the AA are due 
to the hybridization between $f$- and $s$-levels.
It can be seen
in equation (\ref{eq:bandshift})
that in the limit of vanishing hybridization ($V=0$), $B_{-\sigma}=e_{\rm f}$
holds and equations (\ref{maa}) reduce
to the conventional alloy analogy (\ref{aa}). However, for finite
hybridization, $B_{-\sigma}$ and $n_{-\sigma}^{(f)}$ are to be
determined self-consistently by
(\ref{eq:bandshift2}) and (\ref{nf}), respectively, 
possibly providing the alloy energies and concentrations with an
explicit spin-dependence. In each step of the CPA-iteration
$B_{-\sigma}$ and $n_{-\sigma}^{(f)}$ may
change and so does the underlying fictitious
alloy. Equation (\ref{eq:bandshift}) makes
clear that the inclusion of $B_{-\sigma}$ accounts to a certain degree for the
hybridization of $-\sigma$-$f$-electrons with the conduction band and
therefore also to their effective itineracy which
is completely neglected in the
conventional alloy analogy (\ref{aa}).

Before further discussing the MAA, let us remind the reader of the
already mentioned spectral density approximation (SDA). In a previous
work, this theory was derived via mapping of the PAM onto an effective
medium Hubbard model \cite{MNRR98}. Within this effective Hubbard model,
the SDA is justified in the strong coupling limit where the
obtained positions and spectral weights of the lower and upper Hubbard
band coincide with exact results up to the order
$\frac{1}{U}$\cite{HL67}. 
At this point it is noteworthy that exactly the same results as in
reference\cite{MNRR98} can also be obtained by adapting the concept of
the SDA directly onto the PAM which is possible and straightforward from
the above-presented results.
Motivated by the solution of the ``atomic'' limit ($V=0$) we make the
following ansatz 
for the self-energy: 
\begin{equation}
  \label{eq:ansatz_sda}
  \Sigma_{\sigma}(E)= \alpha_{1\sigma} \frac{E-\alpha_{2\sigma}}
{E-\alpha_{3\sigma}}
\end{equation}
The coefficients of this ansatz can now be fitted
in such a way that the high energy expansion of
the self-energy, (\ref{eq:expansionS}) with the coefficients
(\ref{S_coeff}) is fulfilled. One readily arrives at 
\begin{equation}
  \label{eq:sigma_sda}
  \Sigma_{\sigma}^{(\rm SDA)}(E)=\frac {U n_{-\sigma}^{(f)}
    (E-B_{-\sigma}-e_{\rm f})} { E-B_{-\sigma} -e_{\rm f} -U
    (1-n_{-\sigma}^{(f)})} 
\end{equation}
which together with equations (\ref{eq:gf_final}),
(\ref{eq:bandshift2}) and (\ref{nf}) solves the problem. It is
straightforward to show that this is identical to the approximation
proposed in reference \cite{MNRR98}. Therefore, the discussion of the
advantages and disadvantages of the method found therein remains valid.
The main disadvantage was clearly the neglection of quasiparticle damping
in the ansatz~(\ref{eq:ansatz_sda}).

Now turning back to the MAA, its benefits are immediately clear. While
reproducing the high-energy expansion up to the same order as the SDA, it
additionally incorporates quasiparticle damping via the CPA
formalism. And contrary to the conventional alloy analogy, the freedom
of defining the `` alloyed components'' is used to ensure the correct
high-energy expansion~(\ref{eq:expansionS}).

We conclude that the essentials of the qualitatively convincing SDA used
in our previous paper \cite{MNRR98} are incorporated in the MAA and
completed by
a proper quasiparticle damping. So the MAA represents a systematic
improvement and extension of the SDA. Comparison of results for MAA and
SDA will allow to inspect very directly the influence of quasiparticle
damping on magnetic stability in the PAM.

\section{Results and Discussion}
\label{sec:dis}
The presented theory for the periodic Anderson model has been evaluated
for a system characterized by the following parameters: The free
conduction band density of states is chosen to be semi-elliptic with a
width of $W = 1eV$ centred at $E = 0$. The $f$-level is determined by its
distance ($e_{\rm f}$) to the conduction band centre and the intraatomic
Coulomb 
interaction $U$. The latter is $4eV$ throughout the paper. That means we
consider the PAM in the strong coupling regime. Furthermore, we are
mainly interested in the intermediate valence regime, i. e. $e_{\rm f}$ is
presumed to be located within the Bloch-band region ($-0.5eV < e_{\rm f} <
+0.5eV$). It turns out that the physics of the PAM strongly depends on
the total particle density
\begin{equation}
  \label{eq:ntot}
  n^{\rm (tot)}=\sum_{\sigma}(n_{\sigma}^{(f)}+n_{\sigma}^{(s)}); \quad
  0\leq n^{\rm (tot)}\leq4
\end{equation}
We therefore present results for different $n^{\rm (tot)}$, but so that
the upper 
charge excitation at $e_{\rm f} + U$ ("upper Hubbard band") remains in
any case 
unoccupied ($n^{\rm (tot)} < 3$). Another decisive variable is of course
the 
temperature $T$ given in Kelvin. The evaluation of our theory has been done
for a translational symmetric lattice, antiferromagnetic ordering is not
considered. We rather concentrate ourselves on the possibility and the
conditions of spontaneous ferromagnetism.

\begin{figure}[htb]
\begin{center}
    \epsfig{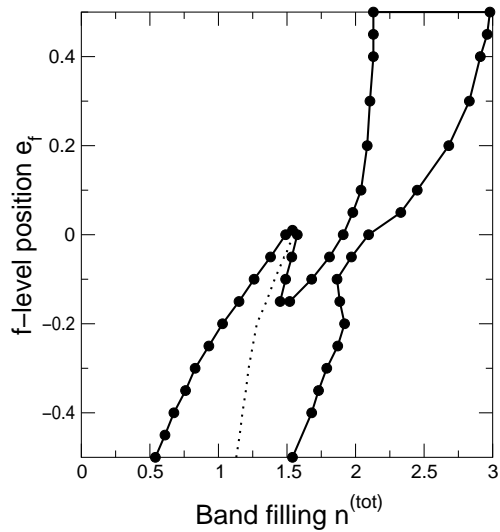}
    \caption{Magnetic phase diagram in the $n^{\rm (tot)}$ - $e_{\rm f}$
      plane for 
      $V=0.1$, $U=4$ and $T=0$.$f-$ and $s-$ magnetizations are parallel
       to the left of the dotted line and anti-parallel to the right.}
\end{center}
\end{figure}
\begin{figure}[htb]
\begin{center}
    \epsfig{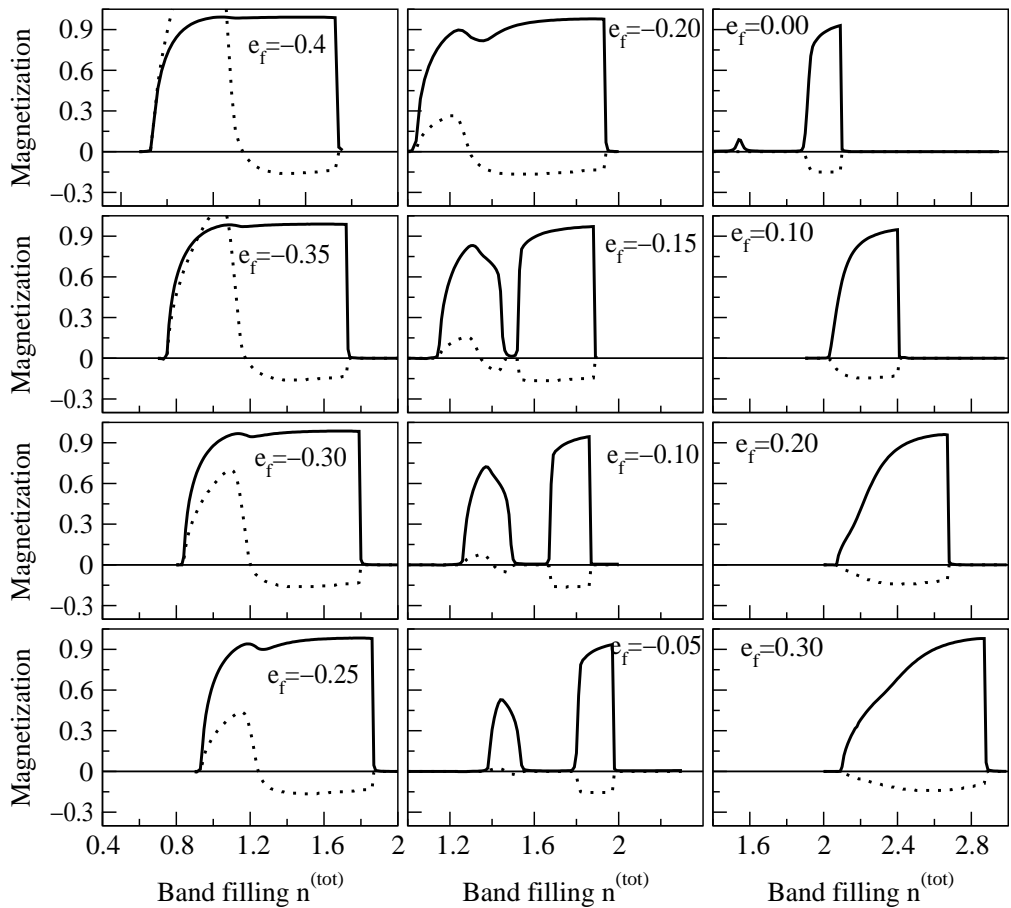}
    \caption{$f-$ (solid line) and $s-$ (dotted line) magnetizations  
     as a function of $n^{\rm (tot)}$ for various values of $e_{\rm f}$. The 
     $s-$magnetization is multiplied by a factor of 5 for better
     visibility. Other parameters are as in figure 1.}
\end{center}
\end{figure}
The magnetic phase diagram in terms of $n^{\rm (tot)}$ and $e_{\rm f}$
for a small 
hybridization $V = 0.1eV$ is plotted in figure 1. For a given $e_{\rm
  f}$ within the 
Bloch band region ferromagnetism becomes possible for particle densities
in between a lower and an upper critical value. These critical values
shift to higher numbers with increasing $f$-level position. Ferromagnetism
is of course basically provoked by the correlated $f$-electrons. Except
for the hybridization with the conduction electrons they are described
by the zero-bandwidth Hubbard model. The latter special case, however,
forbids ferromagnetism. So a finite hybridization $V$ is needed. On the
other hand, one knows that ferromagnetic spin order in the Hubbard model
is bound to further conditions: The band occupation and the Coulomb
correlation $U/W$ must exceed critical values which are different for
different lattice structures \cite{NB89,HN97a,HN99,Hub63}. Furthermore,
the free band should have an assymetric density of
states\cite{Wea98}. SDA and MAA 
both do not
allow ferromagnetism in the "pure" Hubbard model for highly symmetric
free densities of states such as the semi-elliptic one irrespective of
the correlation strength $U/W$. In the PAM with finite hybridization
$V$, however, all these conditions can be met: The 
hybridization creates a finite width of the $f$-dispersion, and the
resulting effective $f$-band turns out to be strongly asymetric,
therewith allowing for a ferromagnetic ground
state. For a given $e_{\rm f}$ the lower critical density in the phase
diagram 
(figure 1) is the analogue to the critical particle density in the
Hubbard 
model \cite{HN97a}. At the upper boundary the lower $f$-charge excitation
is more 
or less filled corresponding to a half-filled band in the strongly
coupled Hubbard model for which antiferromagnetism is to be
expected\cite{And63,HN97a}.

\begin{figure}[htb]
\begin{center}
    \epsfig{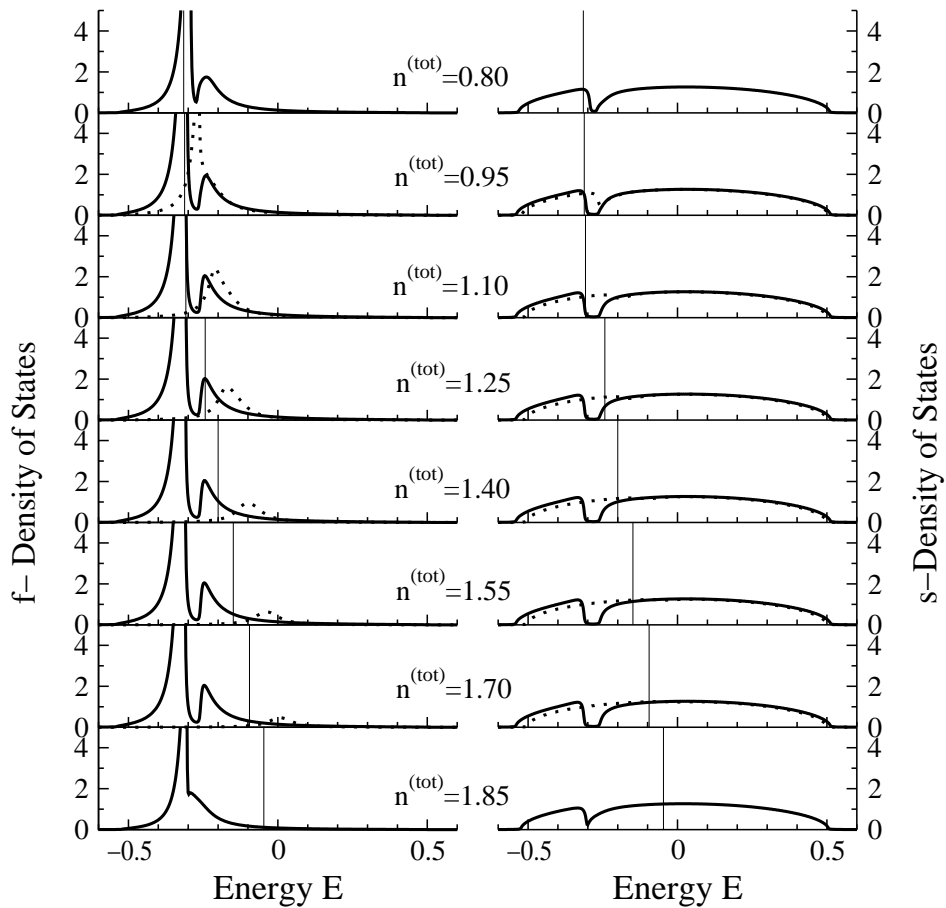}
    \caption{$f-$ and $s-$ quasiparticle densities of states for various
     values of $n^{\rm (tot)}$ and $e_{\rm f}=-0.3$. Full lines for
     spin-up and 
     dotted lines for spin-down. Other parameters are as in figure 1.
     Thin vertical lines show the position of chemical potential.}
\end{center}
\end{figure}
The upper boundary in the phase diagram for a given $e_{\rm f}$ corresponds
always to a first order transition, the lower one to a second order
transition. This can be seen in figure 2 where the spontaneous
$T=0$-magnetization is plotted as function of the total particle density
$n^{\rm (tot)}$. The position of the $f$-level leads to strikingly different
behaviour of the magnetization.

\begin{figure}[htb]
\begin{center}
    \epsfig{file=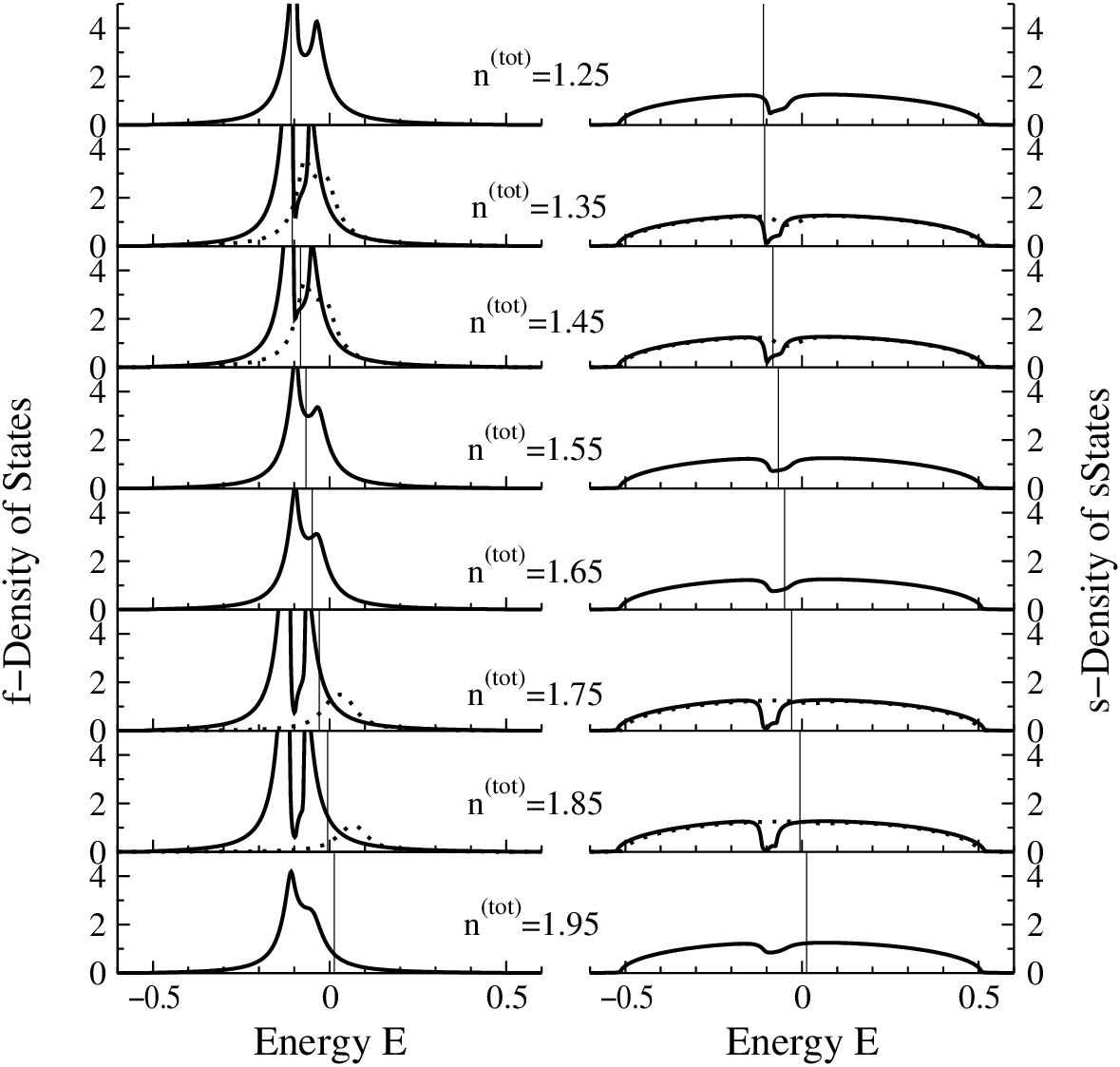,width=12cm,angle=270}
    \caption{Same as figure 3 but $e_{\rm f}=-0.1$.}
\end{center}
\end{figure}

\begin{figure}[htb]
\begin{center}
    \epsfig{file=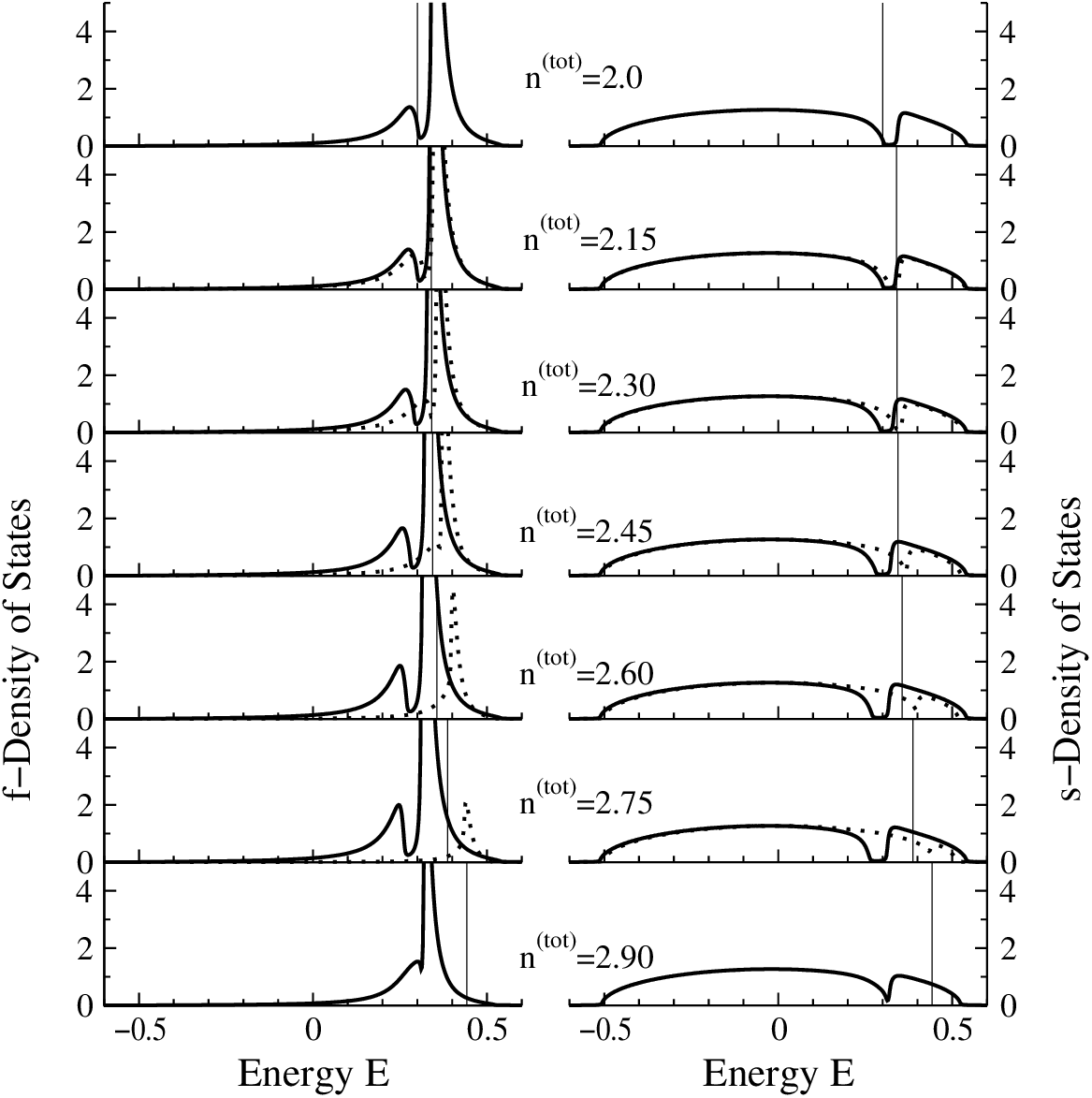,width=12cm, angle=270}
    \caption{Same as figure 3 but $e_{\rm f}=0.3$.}
\end{center}
\end{figure}
With $e_{\rm f}$ in the lower half of the conduction band (left column
in figure 2) 
the $f$-magnetization increases from the lower boundary very rapidly into
saturation in order to perform the mentioned discontinuous transition
into the paramagnetic phase at the upper boundary. Interesting is the
induced $s$-polarization starting positive (ferromagnetic $s$-$f$
coupling) for 
weak electron densities and changing its sign (antiferromagnetic $s$-$f$
coupling) at about $n^{\rm (tot)} = 1.2$. According to the Schrieffer-Wolff
transformation \cite{SW66} at first glance an antiferromagnetic coupling
between 
$s$-and $f$-electrons would be expected. However, the transformation is
bound to the Kondo-limit ($n^{(f)} = 1$) and is not necessarily conclusive
for the intermediate valence region which is investigated here. The
quasiparticle density of states (Q-DOS), plotted in figure 3 for 
$e_{\rm f} = -0.3 eV$ and various electron densities, gives some
evidence how to understand 
the sign change of the $s$-electron polarization. It is instructive to
decompose the Q-DOS into $s$- and $f$-parts, although this decomposition
is artificial because of
the finite hybridization. The latter takes care for the fact that $s$- and
$f$-partial-Q-DOS occupy exactly the same energy regions but of course
with different weights. In the $\uparrow$-parts of both spectra the
hybridization 
gap around $e_{\rm f} = -0.3 eV$ is clearly visible. For practically all
densities $n^{\rm (tot)}$, exhibited in figure 3, the $f$-system is
saturated, 
i. e. there are no down-spin $f$ electrons. It is known from the Hubbard
model \cite{GN88,HN97a} and will be shown in figure
13 for the PAM,
that then the damping of up-spin quasiparticles is
in such a case negligible compared to that of down-spin
quasiparticles. Fine structures
like the hybridization gap are to be seen in the up-spectrum but not in
the down-spectrum. The comparison with the SDA results in ref. \cite{MNRR98}
indeed confirms the interpretation that quasiparticle damping closes the
hybridization gap in the $\downarrow$-spectrum. For the SDA neglects
such damping 
effects, consequently the SDA-$\downarrow$-spectrum, too, exhibits a
gap. There is 
a slight exchange shift in the $s$-part due to hybridization. As long as
the chemical potential $\mu$ is below the hybridization gap the system
contains more up- than down-spin $s$-electrons, the $s$-$f$ coupling is
ferromagnetic. When $\mu$ shifts above the gap (in figure 3 in between 
$n^{\rm (tot)}= 1.10$ and $n^{\rm (tot)} = 1.25$), which exists only in the
up-spectrum, then 
the increase in $n_{\downarrow}^{(s)}$ distinctly exceeds that of
$n_{\uparrow}^{(s)}$, the $s$-$f$ coupling becomes 
antiferromagnetic. That explains the magnetization behaviour in figure 2
(left column) as a density of states effect. Note, however, that the
absolute value of the induced $s$-polarization is always rather small.

\begin{figure}[htb]
\begin{center}
    \epsfig{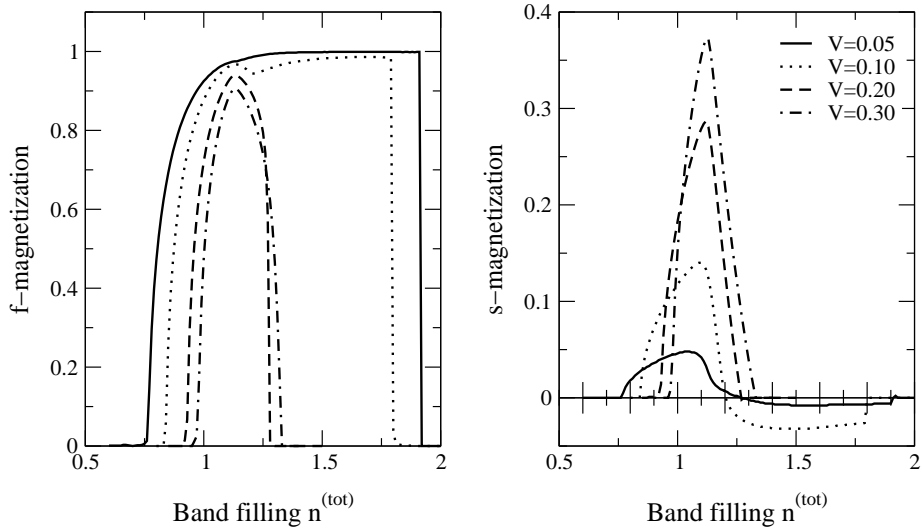}
    \caption{$f$- and $s$- magnetizations as a function of 
     $n^{\rm (tot)}$ for different values of hybridization strength
     $V$ with $e_{\rm f}=-0.3$. Other parameters are as in figure 1.}
\end{center}
\end{figure}
A qualitatively different situation is observed for the magnetization
when $e_{\rm f}$ is near to the centre of the $s$-band (middle column in
figure 2). A reentrant course appears, obviously not connected with the
sign change of the $s$-polarization. It manifests itself in the phase
diagram of figure 1 by the "oscillating" phase boundaries. Again
a look at the respective Q-DOS is quite instructive. figure 4 shows the
example $e_{\rm f} = -0.1 eV$. For such $f$-level position the
hybridization gap 
splits the $f$-dominated part of the spectrum rather symmetrically into
two almost equally weighted peaks. That holds in particular in the
$\uparrow$-spectrum of the ferromagnetic phase 
($n^{\rm (tot)} = 1.35, 1.45, 1.75,1.85$ in figure 4). The chemical
potential 
$\mu$ is located in one of the peaks, 
a situation which according to the simple Stoner criterion favours the
appearance of ferromagnetism. As already discussed for the case of
figure 3 (left column in figure 2) in the ordered phase
$\uparrow$-quasiparticles are
rather long-living while the $\downarrow$-particles are strongly
damped. Again we 
conclude that this is the reason why the hybridization gap exists only
in the $\uparrow$-spectrum. The spontaneous ferromagnetism disappears
when $\mu$ 
enters the hybridization gap in accordance with the Stoner
criterion. The quasiparticles are stronger damped in the paramagnetic
phase than the $\uparrow$-particles in the ordered phase, so that the
hybridization gap is more or less covered by the ``smeared-out''
quasiparticle peaks.

When the bare $f$-level $e_{\rm f}$ has been shifted into the upper half of
the conduction band the polarization type changes once more. The
coupling between $s$- and $f$-electrons is always antiferromagnetic
(right column in figure 2). The Q-DOS in figure 5, concerning the case 
$e_{\rm f} =0.3 eV$, reveals that ferromagnetic order sets in only when $\mu$
is above 
the hybridization gap. The consequence is that there are more
$s$-electrons with spin $\downarrow$ than spin $\uparrow$.

The phase diagram in figure 1 is calculated for a fixed hybridization 
$V =0.1 eV$. It is clear that $V$ has a decisive influence on the
extension of 
the ferromagnetic phase in the $e_{\rm f}$-$n^{\rm (tot)}$ plane. Figure
6 demonstrates 
that the higher the hybridization $V$ the smaller the ferromagnetic
region. On the other hand, the induced polarization of the $s$-electrons
increases with growing hybridization. Antiparallel coupling appears only
for weak $V$. Strong electron fluctuations between $s$-band and $f$-level
generally diminish magnetic stability but enhance the polarization of the
$s$-band. However, the same hybridization is exclusively responsible for
the ferromagnetic order. $V = 0$ does not permit ferromagnetism.

\begin{figure}[htb]
\begin{center}
    \epsfig{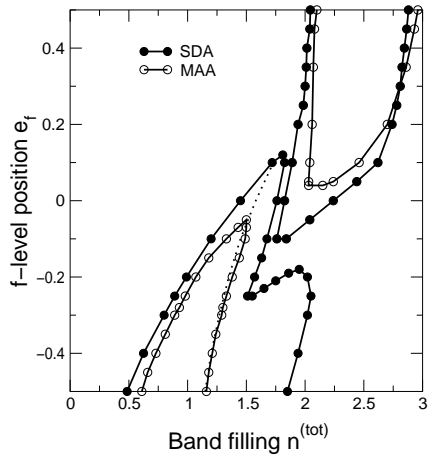}
    \caption{Magnetic phase diagram for MAA and SDA with $V=0.2$, 
     $U=4$ and $T=0$. $f-$ and $s-$ magnetizations are parallel to the left of
     the dotted line and anti-parallel to the right in the SDA. In the
     MAA $f-$ and  
     $s-$ magnetizations are ant-parallel only when $n^{\rm (tot)} > 2$.}
\end{center}
\end{figure}
It is an interesting point to find out to what an extent quasiparticle
damping may influence the possibility and the stability of ferromagnetic
order. This can best be done by a comparison of the results found by
SDA\cite{MNRR98} and by MAA. Both methods are based on the same physical
ideas, MAA
can be classified as "SDA plus quasiparticle damping". Figure 7 shows the
magnetic phase diagram, derived within, respectively, SDA and MAA for
the same set of model parameters. The $e_{\rm f}$-$n^{\rm (tot)}$ area
of the 
ferromagnetic phase is distinctly restricted in the MAA because of the
quasiparticle damping, contrary to the SDA-result which is free of
damping effects. For both methods we find that for lower particle
densities a ferromagnetic, for higher densities an antiferromagnetic
coupling of the $s$-electron spins to the local $f$-moments takes place.
\begin{figure}[htb]
\begin{center}
    \epsfig{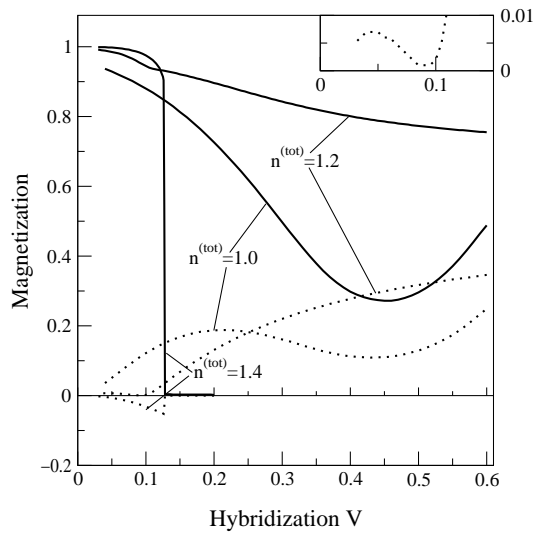}
    \caption{$f-$ (solid line) and $s-$ (dotted line) magnetizations as a 
     function of hybridization strength $V$ for different $n^{\rm (tot)}$
     values for $e_{\rm f} = -0.3$ at $T=0$. The inset shows the $s-$
     magnetization for $n^{\rm (tot)}=1.2$ in the region of small $V$.}
\end{center}
\end{figure}

Up to now we discussed mainly the $e_{\rm f}$- and the 
$n^{\rm (tot)}$-dependence of 
the physical properties of the periodic Anderson model.
Let us now examine the
influence of $V$ in detail. Some results are plotted in
figures 8, 9 and 10 for $e_{\rm f} = -0.3 eV$ and several particle densities
$n^{\rm (tot)}$. The behaviour is not at all unique. For 
$n^{\rm (tot)}= 1.4$ a 
rather weak hybridization $V$ is sufficient to destroy the
ferromagnetism (figure 8). The Curie temperature $T_{\rm c}$ runs through a
maximum (figure 9) pointing therewith to the fact that $V$ provokes two
competing effects. It broadens the $f$-level, thus creating the
precondition for a magnetic order (no ferromagnetism in the
zero-bandwidth Hubbard model \cite{Hub63}!). On the other hand,
increasing $s$-$f$ fluctuations must damage the ferromagnetic order
because the empty $f$-level does not carry a magnetic moment. It is
remarkable that $T_{\rm c}$ goes smoothly to zero (figure 9) although the
$T=0$-magnetization performs a first-order transition (figure 8).
\begin{figure}[htb]
\begin{center}
    \epsfig{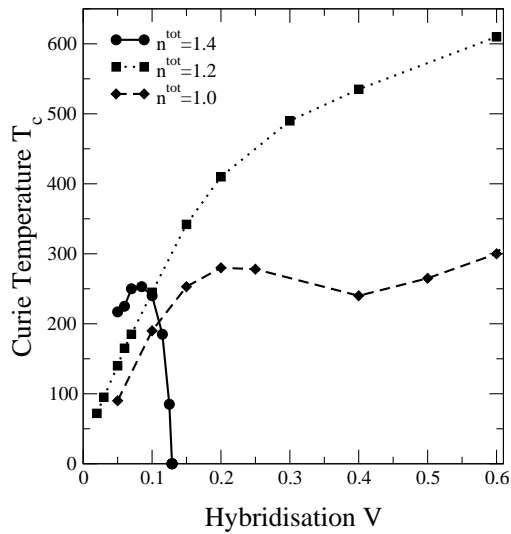}
    \caption{Curie Temperature $T_{\rm c}$ as a function of $V$ for different
      values of $n^{\rm (tot)}$ with $e_{\rm f}=-0.3$.} 
\end{center}
\end{figure}

A completely different $V$-dependence shows up for the band occupations
$n^{\rm (tot)} = 1.0$ and $1.2$. In these cases large hybridization strength
does not destroy the ferromagnetism, but rather enhances
$T_{\rm c}$. By closely examining figure 8, one recognizes a ``kink'' or a
minimum in the respective magnetization curves, for $n^{\rm (tot)}=1.0$
at $V\approx 0.45$ and for $n^{\rm (tot)}=1.2$
at $V\approx 0.09$. By comparing with the densities of states (figure
10), we note that this value approximately coincides with the $V$-value
where $\mu$ enters the hybridization gap in the $\uparrow$-Q-DOS. I.\
e.\ the system becomes a semimetal where only the
$\downarrow$-electrons contribute to 
the electrical current. 
One might speculate that this has something to do with the increasing
$T_{\rm c}$.
However, we cannot
rule out the possiblity that this is an effect of our approximation
rather than an inherent property of the PAM.
\begin{figure}[htb]
\begin{center}
    \epsfig{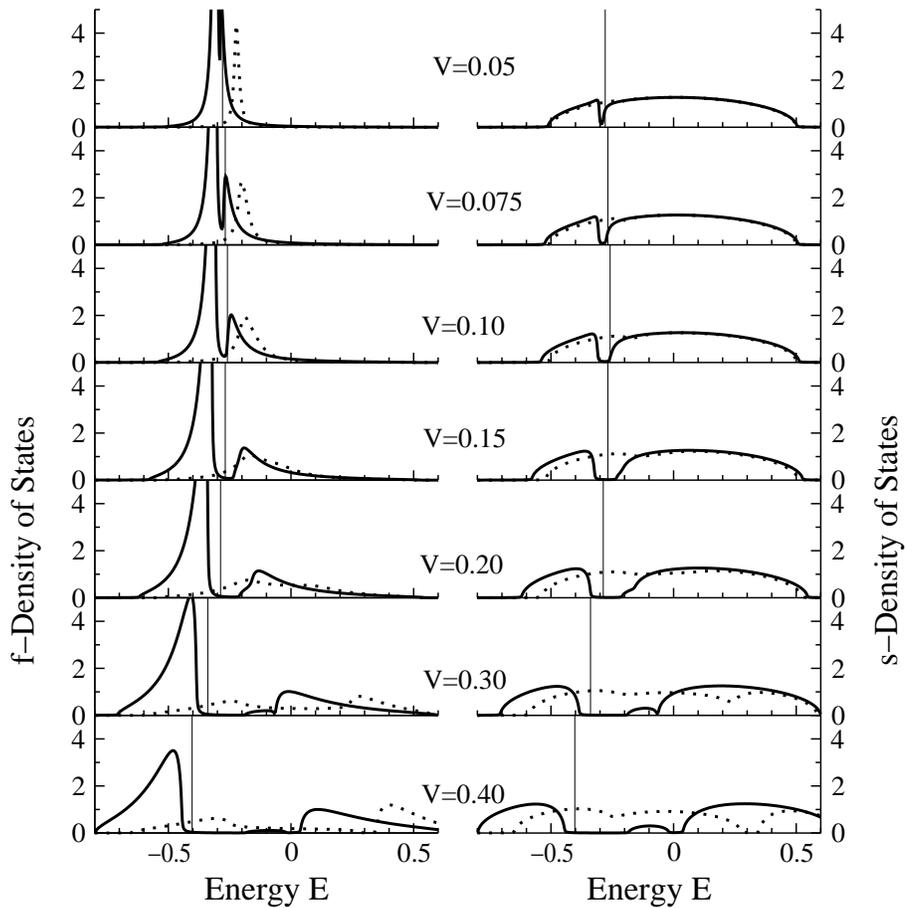}
    \caption{$f$- and $s$- quasiparticle densities of states for various
      values 
      of $V$ with $e_{\rm f}=-0.3$ and $n^{\rm (tot)}=1.2$ at
      $T=0$. Full lines 
      for spin-up and dotted lines for spin-down. Thin vertical lines
      show the position of chemical potential.}
\end{center}
\end{figure}

\begin{figure}[htb]
\begin{center}
    \epsfig{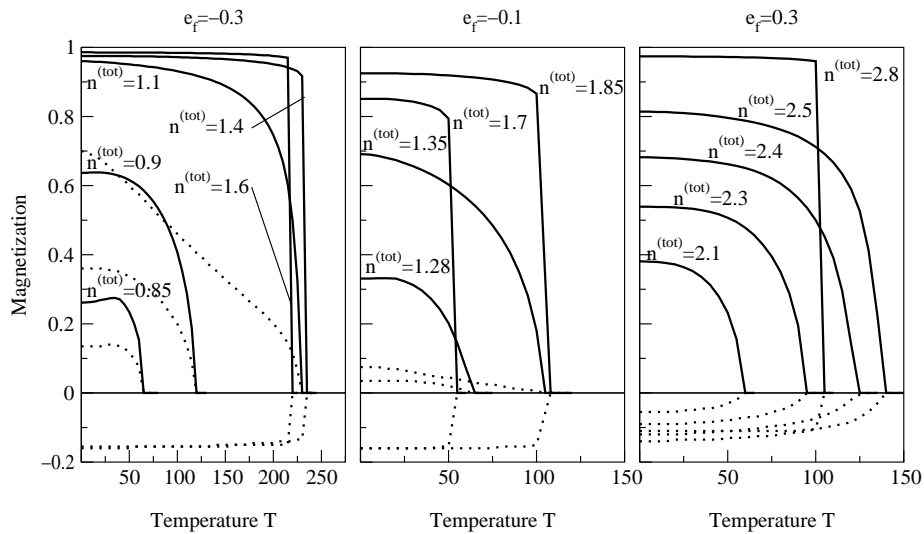}
    \caption{$f-$ (solid line) and $s-$ (dotted line) magnetization as a
     function of temperature for various $e_{\rm f}$ and $n^{\rm (tot)}$ as
     mentioned in the figure. $s-$ mgnetization is multiplied by a
     factor of 5 for better clarity.}
\end{center}
\end{figure}

Let us finally discuss the temperature-dependence of the magnetic
properties as, e. g., the spontaneous magnetization (figure 11). First
order as well as second order transitions appear. Whether first order
transitions are artefacts of our approximate procedure or intrinsic
properties of the PAM is not clear. A similar situation is found when
applying the MAA to the Hubbard model \cite{PHWN98}. In case of a second-order
transition the $f$-magnetization behaves like a Brillouin function,
while the corresponding induced $s$-polarization very often shows
remarkable deviations. The examples plotted in figure 11 demonstrate again
that the $s$-$f$ coupling may be ferromagnetic or antiferromagnet depending
on the band filling.
\begin{figure}[htb]
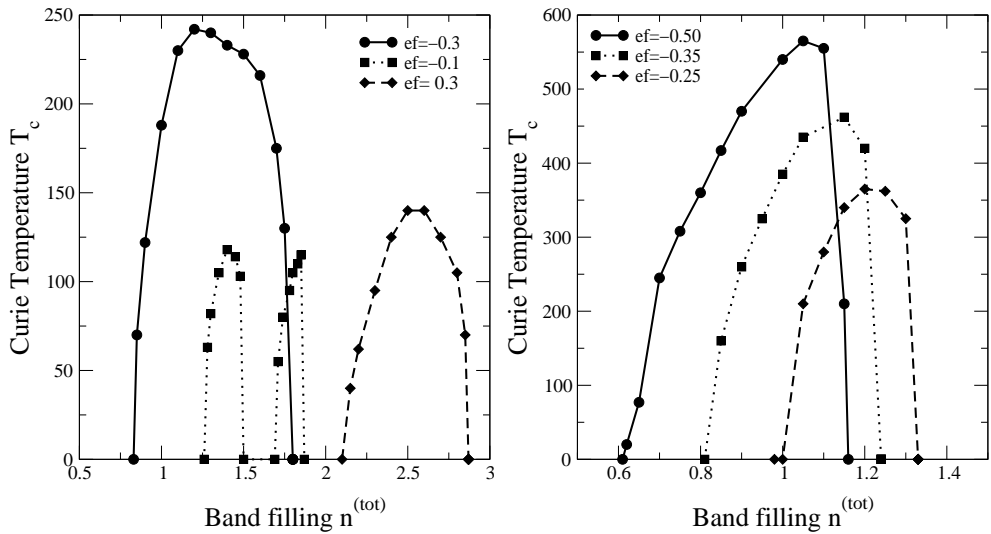

\begin{center}
    \epsfig{file=figures/tc_1.eps, width=7cm, angle=270}
    \epsfig{file=figures/tc_2.eps, width=7cm, angle=270}
     \caption{Curie temperature $T_{\rm c}$ as a function of band
       filling $n^{\rm (tot)}$ 
     for different values of $e_{\rm f}$. The hybridization strength $V=0.1$
     for the left and $V=0.2$ for the right of figure.}
\end{center}
\end{figure}

\begin{figure}[htb]
\begin{center}
    \epsfig{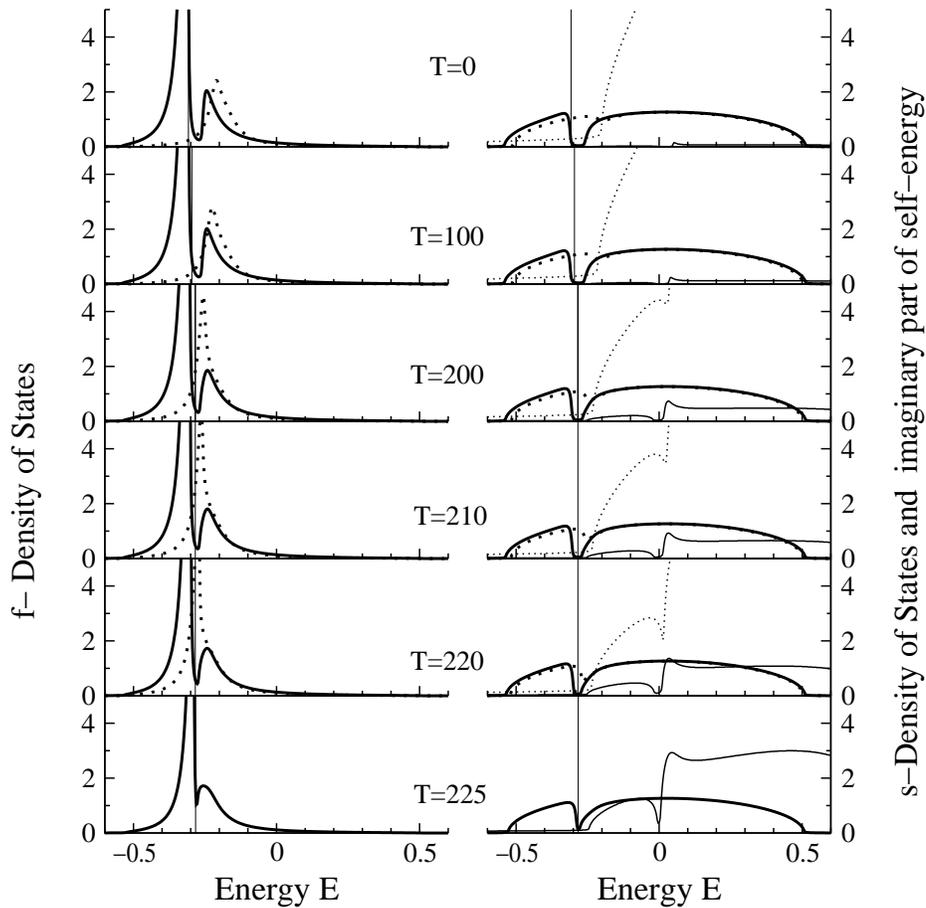}
    \caption{$f$- and $s$- quasiparticle densities of states at various 
     temperatures. $e_{\rm f}$=-0.3, $V$=0.1 and $n^{\rm (tot)}$
     =1.1. Full lines for spin-up 
     and dotted lines for spin-down. Thin vertical lines show the position of 
     chemical potential. Additionally, the imaginary part of the
     self-enery is plotted as thin lines in the right column. For better
     visibility, it is multiplied by $-100$.}
\end{center}
\end{figure}
A key-quantity of ferromagnetism is the Curie-temperature $T_{\rm c}$,
which is 
of course decisively influenced by the bandfilling (figure 12). The
$n^{\rm (tot)}$-dependence is very much more regular than the $V$-dependence
exhibited in figure 9. The reentrant behaviour for $e_{\rm f} = -0.1 eV$
corresponds to that of the $T = 0$-moment in figure 2 (middle column). The
calculated $T_{\rm c}$-values are of a realistic order of magnitude. The
transition into the paramagnetic phase ($T_{\rm c} = 0$) seems to be always
continuous even if the break-down of the $T = 0$-moment is discontinuous
(figure 2).  The temperature-dependence of the magnetization is due to a
respective behaviour of the Q-DOS (figure 13). There is a distinct
spin asymmetry in the lower $f$-like peak ("lower Hubbard band") which causes
the spontaneous magnetic moment. The up-spin part is characterized by a
hybridization gap, which is not visible in the down-spin spectrum
(cf. figures 3, 4 and 5). As already mentioned, this is due to
quasiparticle damping, which for low temperatures is very much 
stronger for $\downarrow$- than for $\uparrow$-quasiparticles
(cf. figure 13). 
With
increasing temperature 
the damping of the $\uparrow$-particles, is getting larger and
that of the $\downarrow$ particles smaller.
For increasing temperatures, a dip develops in the spin-$\downarrow$
Q-DOS, which eventually merges with the spin-$\uparrow$ hybridization
gap at $T=T_{\rm c}$.
The
spin asymmetry 
is then removed. The induced spin polarization of the conduction band is
always very weak so that the assumption that the collective order is
based on an RKKY-like coupling via the polarized conduction electrons
appears unlikely. It is rather to believe that the observed
ferromagnetism is due to strong electron correlations in the narrow
"$f$-band" as it happens in the single-band Hubbard model
\cite{HN96,HN97b}. The
coupling to the $s$-band via $V$ takes care for the finite width of the
original $f$-level being in this sense the basic precondition for
ferromagnetic order in the periodic Anderson model.

\section{Conclusion}
\label{sec:conclu}
A "modified alloy analogy" (MAA) \cite{HN96,PHWN98}, previously
introduced as an approach to the single-band Hubbard model \cite{Hub63}
has been
applied to the periodic Anderson model (PAM). A high-energy expansion of
relevant Green functions as well as a corresponding expansion of the
determination equation for the CPA-self-energy have been used to find the
optimum alloy analogy for the PAM. This alloy analogy is then used to
solve the PAM-many body problem within the CPA. By construction the MAA
represents a strong-coupling approach, therewith most probably suitable
to describe spontaneous ferromagnetism in the PAM. It can be considered
as an extension and improvement of the "spectral density approach" (SDA)
\cite{GN88,NB89,HN97a,PHWN98}, mainly by inclusion of quasiparticle
damping. It 
cannot reproduce, however, the low-energy features of the PAM (Kondo
resonance)\cite{hewson,Jar95}, being certainly not so decisive with
respect to magnetic 
stability. It incorporates, however, important higher correlation
functions ("spin-dependent band shift") which guarantee the correct
strong-coupling behaviour \cite{HL67}.

The present study focuses on the possibility and the stability of
ferromagnetism in the PAM: Magnetic phase diagrams are constructed in
terms of relevant model parameters such as hybridization strength $V$,
$f$-level position $e_{\rm f}$ and total particle density 
$n^{\rm (tot)}$. For this 
work we are mainly interested in the intermediate valence regime,
i. e. that $e_{\rm f}$ is chosen within the energy region of the "free"
Bloch 
band. The same holds for the chemical potential $\mu$. As usual "$f$-level"
and conduction band, respectively, are assumed as non-degenerate. The
intraatomic Coulomb interaction $U$ of the $f$ electrons leads to a
splitting of $e_{\rm f}$ into two sublevels at $e_{\rm f}$ and 
$e_{\rm f} + U$. $U$ is chosen 
so that the upper charge excitation remains unoccupied i. e. $0 <
n^{\rm (tot)} < 3$. The influence of $V$ is multifold. First of all it
provokes 
a finite width of the lower $f$-quasiparticle peak that turns out a
basic prerequisite for ferromagnetism in the intermediate valence
PAM. On the other hand, too strong $s$-$f$ fluctuations due to $V$
destabilize 
the ferromagnetic order. Additionally $V$ provokes a hybridization gap in
the energy spectrum, which, however, is getting closed by too strong
quasiparticle damping. If the $f$-magnetization is almost saturated then
the $\uparrow$-spectrum shows a hybridization gap, the
$\downarrow$-spectrum not. This has 
some influence on the induced $s$-polarization which can be parallel or
antiparallel to the $f$-moment. The derived Curie temperatures are of
reasonable orders of magnitude. All magnetic properties of PAM can be
illustratively traced back and reasoned by inspection of the respective
quasiparticle density of states.

\ack
We wish to thank T. Herrmann for very helpful discussions. Financial
support by the \textit{Volkswagen-Foundation} within the project
\textit{``Phasendiagramm des Kondo-Gitter-Modells''} is gratefully
acknowledged. One of us (D.\ M.\ ) wants to thank the
\textit{Friedrich-Naumann-Foundation} for supporting his work.
\end{document}